\documentstyle[12pt,epsfig]{article}
\def\lbldef#1#2{\expandafter\gdef\csname #1\endcsname {#2}}

\def\href#1#2{#2}  

\begin{document}
\baselineskip=15.5pt
\pagestyle{plain}
\setcounter{page}{1}

\begin{titlepage}

\begin{flushright}
IC/2000/189\\
hep-th/0012246
\end{flushright}
\vspace{10 mm}

\begin{center}
{\Large Cosmology of Dilatonic Brane World}

\vspace{5mm}

\end{center}

\vspace{5 mm}

\begin{center}
{\large Donam Youm\footnote{E-mail: youmd@ictp.trieste.it}}

\vspace{3mm}

ICTP, Strada Costiera 11, 34014 Trieste, Italy

\end{center}

\vspace{1cm}

\begin{center}
{\large Abstract}
\end{center}

\noindent

We study cosmological solutions in the dilatonic brane world models.  
The effective four-dimensional equations on the brane are analyzed for 
the models with one positive tension brane and two branes with 
tensions of opposite signs.  Just as in the non-dilatonic brane case, 
the conventional Friedmann equations of the four-dimensional universe 
are reproduced to the leading order in matter energy density for the 
model with one brane and the introduction of a radion potential is 
required in order to reproduce the Friedmann equations with the correct 
sign for the model with two branes.  

\vspace{1cm}
\begin{flushleft}
December, 2000
\end{flushleft}
\end{titlepage}
\newpage

Over the past couple of years, much attention has been paid to theories 
with extra dimensions \cite{add,aadd,add2,rs1,rs2,rs3}, as such theories 
open up the possibility of solving the hierarchy problem in particle physics 
and make the string scale and the extra dimensions accessible to the future 
accelerators.  Such theories assume that fields of the Standard Model are 
confined to a three-brane embedded in higher-dimensional spacetime, whereas 
gravity can propagate in the bulk.  In the scenario proposed by Randall 
and Sundrum (RS) \cite{rs1,rs2,rs3}, four-dimensional gravity with a 
negligible correction is reproduced even with an infinitely large extra 
dimension, since the gravity is effectively localized on the brane.  

Lots of effort has been made to understand cosmology in brane models. 
Initially, it was observed \cite{bdl} that brane models (with vanishing 
bulk cosmological constant $\Lambda$ and brane tension $\sigma$) have 
non-conventional cosmological solutions where the Hubble parameter $H$ is 
proportional to the energy density $\varrho$ of matter on the brane, whereas 
in standard conventional cosmology $H\propto\varrho$.  This problem was 
resolved \cite{cgkt,cgs,bdel,ftw} by considering the RS brane world model, 
i.e. by setting $\Lambda$ and $\sigma$ to be nonzero.  By assuming that 
$\varrho\ll\sigma$, one reproduces the conventional cosmology to the leading 
order in $\varrho$ for the RS model with one brane and an infinitely large 
extra dimension (called the RS2 model).  However, for the RS model with two 
branes (called the RS1 model), the cosmological solution on the visible 
brane (the TeV brane) has the wrong-signed Friedmann equations.  
Furthermore, the matter energy densities $\varrho$ and $\varrho_*$ on the 
invisible and the visible branes are correlated to have the opposite signs, 
which leads to the unphysical result that the matter energy density on 
either of the branes has to be negative.  As was speculated in Refs. 
\cite{gw2,bdl,cgkt} and later explicitly shown in Ref. \cite{cgrt} (see also 
Ref. \cite{cf} for the brane world cosmology with a radion stabilizing 
potential), such undesirable results are resolved by including a radion 
stabilizing potential \cite{gw1,gw2,gw3}.  In the limit of very heavy radion 
mass, one of the equations of motion which correlated the energy densities 
on the two branes disappears and the standard cosmology is reproduced.  
Another approach to resolve the problem of the brane cosmology emphasized in 
Ref. \cite{bdl} was proposed in Refs. \cite{kkop1,kkop2}, where it is shown 
that the standard Friedmann equations can be reproduced (without any 
approximation) by introducing the nonzero extra-spatial component of the bulk 
energy-momentum tensor (even without the bulk cosmological constant and 
the brane tension).  Such nonzero bulk energy-momentum tensor component acts 
to stabilize the radius of the extra spatial dimension even in the 
absence of the second brane.  

It was found out \cite{youm1,youm2} that the RS type brane world scenario can 
be extended to the dilatonic branes, since gravity can also be localized 
on the dilatonic branes when the warp factor decreases, for which 
case the tension of the brane is positive.  Dilatonic domain walls 
appear in string theories quite often when (intersecting) branes in 
ten or eleven dimensions are compactified.  It is the purpose of this 
note to study cosmological solutions in the dilatonic brane world.  
(The previous works on the brane world cosmology with the bulk dilatonic 
or Brans-Dicke scalar can be found in Refs. \cite{mw,mb,mm,bv}.) 

We consider the following action:
\begin{equation}
S=\int d^5x\sqrt{-\hat{g}}\left[{1\over{2\kappa^2_5}}{\cal R}-
{4\over 3}\partial_M\phi\partial^M\phi-e^{-2\alpha\phi}\Lambda\right]
+\int d^4x\sqrt{-g}\left[{\cal L}_{mat}-\sigma e^{-\alpha\phi_0}\right],
\label{action}
\end{equation}
where $\sigma$ is the tension of the brane located at $y=0$, $\phi_0\equiv
\phi|_{y=0}$,  
and ${\cal L}_{mat}$ is the Lagrangian density for all the matter confined 
on the three-brane.  

When ${\cal L}_{mat}=0$, the equations of motion following from the action 
(\ref{action}) admit the following static three-brane solution \cite{youm1}:
\begin{eqnarray}
g_{MN}dx^Mdx^N={\cal W}\left[-dt^2+(dx^1)^2+(dx^2)^2+(dx^3)^2\right]+dy^2,
\cr
\phi={1\over\alpha}\ln(1-K|y|)+\phi_0,\ \ \ \ \ \ \ 
{\cal W}=(1-K|y|)^{{16\kappa^2_5}\over{9\alpha^2}},
\label{statdw}
\end{eqnarray}
where the coefficient $K$ is determined by $\Lambda$ in the following way:
\begin{equation}
K={{3\alpha^2}\over 2}e^{-\alpha\phi_0}\sqrt{{3\Lambda}\over{9\alpha^2-
32\kappa^2_5}},
\label{kcoeff}
\end{equation}
and $\phi_0$ is an arbitrary constant.  
The following fine-tuned value of $\sigma$ in terms of $\Lambda$ is 
fixed by the boundary condition at $y=0$:
\begin{equation}
\sigma=8\sqrt{{3\Lambda}\over{9\alpha^2-32\kappa^2_5}}.
\label{brntnsn}
\end{equation}
It was shown \cite{youm1,youm3,youm2} that as long as $\sigma$ takes 
positive value given by Eq. (\ref{brntnsn}) (i.e. the case with the 
${\bf Z}_2$-symmetry and naked singularities on both sides of the brane) 
there exists normalizable Kaluza-Klein zero mode for the bulk graviton.  
Thereby, the RS type brane world scenario can be extended to the dilatonic 
domain wall case.  

When ${\cal L}_{mat}\neq 0$, even if $\sigma$ takes the fine tuned value 
(\ref{brntnsn}), the brane world becomes no longer static, i.e. the brane 
world undergoes cosmological expansion.  We are interested in the 
cosmological model for which the principle of homogeneity and isotropy in 
the three-dimensional space of the brane universe is satisfied.  On the 
other hand, the presence of the brane breaks the isometry along the extra 
spatial direction.  The general form of the metric Ansatz satisfying these 
requirements is
\begin{equation}
\hat{g}_{MN}dx^Mdx^N=-n^2(t,y)dt^2+a^2(t,y)\gamma_{ij}dx^idx^j+b^2(t,y)dy^2.
\label{metans}
\end{equation}
Here, $\gamma_{ij}$ is the maximally symmetric three-dimensional metric 
given in the Cartesian and spherical coordinates by
\begin{equation}
\gamma_{ij}dx^idx^j=\left(1+{k\over 4}\delta_{mn}x^mx^n\right)^{-2}
\delta_{ij}dx^idx^j
={{dr^2}\over{1-kr^2}}+r^2(d\theta^2+\sin^2\theta d\phi^2),
\label{threemet}
\end{equation}
where $k=-1,0,1$ respectively for the three-dimensional space with 
the negative, zero and positive spatial curvature.  

By varying the action (\ref{action}) with respect to $\hat{g}_{MN}$, one 
obtains the Einstein's equations
\begin{equation}
{\cal G}_{MN}=\kappa^2_5T_{MN},
\label{eineqs}
\end{equation}
with the energy-momentum tensor given by
\begin{eqnarray}
T_{MN}&=&-{4\over 3}\hat{g}_{MN}\partial_P\phi\partial^P\phi+{8\over 3}
\partial_M\phi\partial_N\phi-\hat{g}_{MN}e^{-2\alpha\phi}\Lambda
\cr
& &-{\sigma\over b}e^{-\alpha\phi}\delta(y)\delta^{\mu}_M\delta^{\nu}_N
g_{\mu\nu}+{1\over b}\delta(y)\delta^{\mu}_M\delta^{\nu}_NT^{mat}_{\mu\nu},
\label{emtens}
\end{eqnarray}
where $T^{mat}_{\mu\nu}=-{2\over\sqrt{-g}}{{\delta(\sqrt{-g}{\cal L}_{mat})}
\over{\delta g^{\mu\nu}}}$ is the energy-momentum tensor of the 
matter fields on the three-brane.  Since we wish to model the matter in 
the brane universe by a perfect fluid, $T^{mat}_{\mu\nu}$ in the comoving 
coordinates takes the following form:
\begin{equation}
T^{mat\,\mu}_{\ \ \ \ \ \ \nu}={\rm diag}(-\varrho,\wp,\wp,\wp),
\label{pfemten}
\end{equation}
where $\varrho$ and $\wp$ are the energy density and pressure of  
matter on the three-brane as measure in the rest frame.  
So, the Einstein's equations (\ref{eineqs}) take the following form:
\begin{eqnarray}
{3\over n^2}{\dot{a}\over a}\left({\dot{a}\over a}+{\dot{b}\over b}\right)-
{3\over b^2}\left[{a^{\prime}\over a}\left({a^{\prime}\over a}-{b^{\prime}
\over b}\right)+{a^{\prime\prime}\over a}\right]+{{3k}\over a^2}=
\ \ \ \ \ \ \ \ \ \ \ \ \ \ \ \ \ \ 
\cr
\kappa^2_5\left[{4\over 3}n^{-2}\dot{\phi}^2+{4\over 3}b^{-2}\phi^{\prime\,2}
+e^{-2\alpha\phi}\Lambda+(\sigma e^{-\alpha\phi}+\varrho)
{{\delta(y)}\over b}\right],
\label{expeineqs1}
\end{eqnarray}
\begin{eqnarray}
{1\over b^2}\left[{a^{\prime}\over a}\left(2{n^{\prime}\over n}+{a^{\prime}
\over a}\right)-{b^{\prime}\over b}\left({n^{\prime}\over n}+2{a^{\prime}
\over a}\right)+2{a^{\prime\prime}\over a}+{n^{\prime\prime}\over n}\right]
\ \ \ \ \ \ \ \ \ \ \ \ \ \ \ \ \ \ \ \ \ \ \ \ \ \ 
\cr
+{1\over n^2}\left[{\dot{a}\over a}\left(2{\dot{n}\over n}
-{\dot{a}\over a}\right)+{\dot{b}\over b}\left({\dot{n}\over n}
-2{\dot{a}\over a}\right)-2{\ddot{a}\over a}-{\ddot{b}\over b}\right]
-{k\over a^2}=\ \ \ \ \ 
\cr
\kappa^2_5\left[{4\over 3}n^{-2}\dot{\phi}^2-{4\over 3}b^{-2}\phi^{\prime\, 2}
-e^{-2\alpha\phi}\Lambda+(\wp-\sigma e^{-\alpha\phi}){{\delta(y)}\over b}
\right],
\label{expeineqs2}
\end{eqnarray}
\begin{equation}
{n^{\prime}\over n}{\dot{a}\over a}+{a^{\prime}\over a}{\dot{b}\over b}
-{\dot{a}^{\prime}\over a}={8\over 9}\kappa^2_5\dot{\phi}\phi^{\prime},
\label{expeineqs3}
\end{equation}
\begin{equation}
{3\over b^2}{a^{\prime}\over a}\left({a^{\prime}\over a}+{n^{\prime}\over n}
\right)-{3\over n^2}\left[{\dot{a}\over a}\left({\dot{a}\over a}-{\dot{n}
\over n}\right)+{\ddot{a}\over a}\right]-{{3k}\over 
a^2}=\kappa^2_5\left[{4\over 3}n^{-2}\dot{\phi}^2+{4\over 3}b^{-2}
\phi^{\prime\,2}-e^{-2\alpha\phi}\Lambda\right],
\label{expeineqs4}
\end{equation}
where the overdot and the prime respectively denote derivatives w.r.t. 
$t$ and $y$.

With the assumption of homogeneity and isotropy on the brane world, the 
dilaton $\phi$ does not depend on the spatial coordinates $x^i$ ($i=1,2,3$) 
of the three-brane.  So, the equation of motion for the dilaton takes the 
following form:
\begin{eqnarray}
{8\over 3}{1\over n^2}\left[\ddot{\phi}-\dot{\phi}\left({\dot{n}\over n}
-3{\dot{a}\over a}-{\dot{b}\over b}\right)\right]-{8\over 3}{1\over b^2}
\left[\phi^{\prime\prime}+\phi^{\prime}\left({n^{\prime}\over n}
+3{a^{\prime}\over a}-{b^{\prime}\over b}\right)\right]
\cr
=2\alpha\Lambda e^{-2\alpha\phi}+\alpha\sigma e^{-\alpha\phi}
{{\delta(y)}\over b}.\ \ \ \ \ \ \ \ \ \ \ \ 
\label{eqndil}
\end{eqnarray}

We assume that a solution to the above equations of motion is continuous 
everywhere, especially across $y=0$, where the brane is located, in order 
to have a well-defined geometry.  However, its derivatives w.r.t. $y$ 
are discontinuous at $y=0$ due to the $\delta$-function like brane 
source there.  The following boundary conditions on the first derivatives 
of the metric components at $y=0$ are obtained by integrating Eqs. 
(\ref{expeineqs1}) and (\ref{expeineqs2}) over the infinitesimal interval 
around $y=0$ w.r.t. $y$:
\begin{equation}
{{[a^{\prime}]_0}\over{a_0b_0}}=-{\kappa^2_5\over 3}(\sigma
e^{-\alpha\phi_0}+\varrho),
\label{bc1}
\end{equation}
\begin{equation}
{{[n^{\prime}]_0}\over{n_0b_0}}=-{\kappa^2_5\over 3}(\sigma
e^{-\alpha\phi_0}-3\wp-2\varrho),
\label{bc2}
\end{equation}
where the subscript $0$ denotes quantities evaluated at $y=0$, e.g. $a_0(t)
\equiv a(t,0)$, and $[F]_0\equiv F(0^+)-F(0^-)$ denotes the jump of $F(y)$ 
across $y=0$.  Similarly, from the dilaton equation (\ref{eqndil}), one 
obtains the following boundary condition on the first derivative of $\phi$ 
at $y=0$:
\begin{equation}
{{[\phi^{\prime}]_0}\over{b_0}}=-{3\over 8}\alpha\sigma e^{-\alpha\phi_0}.
\label{dilbc}
\end{equation}

The effective four-dimensional equations of motion on the three-brane 
can be obtained \cite{bdl} by taking the jumps and the mean values 
of the above five-dimensional equations of motion across $y=0$ and 
then applying the boundary conditions (\ref{bc1}-\ref{dilbc}) 
on the first derivatives.  Here, the mean value of a function $F$ across 
$y=0$ is defined as $\sharp F\sharp\equiv [F(0^+)+F(0^-)]/2$.  In this paper, 
we consider solutions invariant under the ${\bf Z}_2$ symmetry, $y\to -y$, 
i.e. the solutions depending on $y$ through $|y|$.  Then, the mean values of 
the first derivatives across $y=0$ vanish.  We also note that it is 
always possible to choose a gauge so that $n_0(t)\equiv n(t,0)$ is constant 
without introducing the cross term $\hat{g}_{04}$.  Making use of this fact, 
we scale the time coordinate $t$ to be the cosmic time for the brane universe, 
namely $n_0=1$.  

First, by taking the jump of the $(0,4)$-component Einstein equation 
(\ref{expeineqs3}), one obtains the following conservation of energy 
equation for the brane universe with the scale factor $a_0$:
\begin{equation}
\dot{\varrho}+3(\wp+\varrho){\dot{a}_0\over a_0}=0.
\label{cseneq}
\end{equation}
So, despite the energy flow along the $y$-direction, as indicated in Eq. 
(\ref{expeineqs3}), the energy conservation law in the brane universe takes 
the conventional form of the standard four-dimensional universe.  A 
consequence for this fact is that for the brane matter satisfying the 
equation of state of the form $\wp=w\varrho$ with a constant $w$, the 
dependence of $\varrho$ on $a_0$ has the usual form $\varrho\propto 
a^{-3(1+w)}_0$.

Next, by taking the mean value of the $(4,4)$-component Einstein equation 
(\ref{expeineqs4}) across $y=0$, one obtains the following Friedmann-type 
equation of the brane universe:
\begin{eqnarray}
{\dot{a}^2_0\over a^2_0}+{\ddot{a}_0\over a_0}+{k\over a^2_0}&=&
{\kappa^4_5\over{36}}(\varrho-3\wp)\sigma e^{-\alpha\phi_0}+
\left({\kappa^4_5\over{18}}\sigma^2-{\kappa^2_5\over{64}}\alpha^2\sigma^2
+{\kappa^2_5\over 3}\Lambda\right)e^{-2\alpha\phi_0}
\cr
& &-{\kappa^4_5\over{36}}\varrho(\varrho+3\wp)
-{4\over 9}\kappa^2_5\dot{\phi}^2_0.
\label{friedeq}
\end{eqnarray}
The $\sim e^{-2\alpha\phi_0}$ term on the RHS vanishes when $\sigma$ takes 
the fine-tuned value (\ref{brntnsn}).  When $\sigma$ is bigger [smaller] 
than the fine-tuned value, the effective cosmological constant term behaving 
as $\sim e^{-2\alpha\phi_0}$ is positive [negative].  However, unlike the 
non-dilatonic brane world case, we have additional negative contribution 
(the last term on the RHS) to the cosmological constant from the varying 
dilaton field with $t$.  This can be understood from Eq. (\ref{expeineqs3}), 
which indicates the flow of the dilaton energy along the $y$-direction.  
The extremely small positive cosmological constant observed in our universe 
restricts the dilaton to vary very slowly on the brane or to be stabilized 
due to some mechanism.  When $\sigma$ does not take the fine-tuned value, 
the varying $\phi_0$ implies the varying cosmological constant in the 
brane universe with $t$.  If we assume $\varrho\ll\sigma$, then to the 
leading order in $\varrho$ we have $H^2\propto\varrho$ as in conventional 
cosmology.  However, unlike the non-dilatonic brane world case, the 
coefficient of the $\sim\varrho$ term varies with $t$, if $\dot{\phi}_0
\neq 0$.  The $\Lambda=0$ case corresponds to cosmology in the self-tuning 
brane world \cite{adks,kss}.  In this case, the four-dimensional 
effective cosmological constant term ($\sim e^{-2\alpha\phi_0}$) is 
nonzero for a general value of $\alpha$, as was previously observed 
\cite{youm3,flln} in the four-dimensional effective action.  This cosmological 
constant term vanishes when $\alpha^2=32\kappa^2_5/9$.  Contrary to the result 
in Ref. \cite{flln}, this nonzero cosmological constant cannot be 
cancelled by introducing another brane with the fine-tuned tension, 
since the RHS of Eq. (\ref{friedeq}) generally does not receive 
contribution from another brane at different $y$.  This difference may be 
attributed to the fact that the Friedmann-like equation (\ref{friedeq}) 
contains information local in the $y$-direction (i.e. only from $y=0$), 
whereas the four-dimensional effective action contains contribution from 
all the possible values of $y$.  Note, so far, we have not assumed the 
radius $b$ of the extra dimension to be constant.  

We now discuss an approximate solution to the equations of motion.  
We assume the radius of the extra space to be stable, i.e. $b=b_0=const$, 
even in the absence of a stabilizing potential.  We assume that the brane 
matter satisfies the equation of state of the form $\wp=w\varrho$ with 
a constant $w$.  The generalization of the static brane solution 
(\ref{statdw}) can be parameterized as
\begin{eqnarray}
a(t,y)&=&a_0(t)\left[1+A(t)|y|+A_2(t)y^2+\cdots
\right]^{{8\kappa^2_5}\over{9\alpha^2}},
\cr
n(t,y)&=&\left[1+N(t)|y|+N_2(t)y^2+\cdots\right]^{{8\kappa^2_5}\over
{9\alpha^2}},
\cr
\phi(t,y)&=&{1\over\alpha}\ln\left[1+\Phi(t)|y|+\Phi_2(t)y^2+\cdots\right]
+\phi_0(t).
\label{expbrnpara}
\end{eqnarray}
The coefficients of the $|y|$ terms can be determined by applying the 
boundary conditions (\ref{bc1}-\ref{dilbc}) on the first derivatives.  
The resulting expressions are
\begin{equation}
A=-Kb_0-{{3\alpha^2}\over{16}}\varrho b_0,\ \ \ 
N=-Kb_0+{{3\alpha^2}\over{16}}(3w+2)\varrho b_0,\ \ \ 
\Phi=-Kb_0.
\label{fstcoeffs}
\end{equation}
Note, in the presence of matter fields on the brane, $K={{3\alpha^2}\over{16}}
e^{-\alpha\phi_0}\sigma$ changes with time, since $\phi_0=\phi(t,0)$ is 
in general a function of $t$.  So, the time dependence of the above 
coefficients comes not only from $\varrho(t)$ but also from $K$.  Of course, 
when there are no matter fields and $\sigma$ takes the fine-tuned value 
(\ref{brntnsn}), $K$ is a constant and the above inflationary brane 
solution reduces to the static brane solution (\ref{statdw}).  The linear 
order (in $|y|$) part of the above non-static brane solution 
(\ref{expbrnpara}) suggests the following approximate solution, valid for 
any $y$, to the first order in $\varrho$:
\begin{eqnarray}
a(t,y)&\approx&a_0(t)(1-Kb_0|y|)^{{8\kappa^2_5}\over{9\alpha^2}}
[1+\varrho(t)f(t,y)],
\cr
n(t,y)&\approx&(1-Kb_0|y|)^{{8\kappa^2_5}\over{9\alpha^2}}
[1+\varrho(t)g(t,y)],
\cr
\phi(t,y)&\approx&{1\over\alpha}\ln(1-Kb_0|y|)+\phi_0(t),
\label{fstrhosol}
\end{eqnarray}
where
\begin{eqnarray}
f(t,y)&=&-{{3\alpha^2}\over{32K}}\left[(1-Kb_0|y|)^{-{{16\kappa^2_5}\over
{9\alpha^2}}}-1\right],
\cr
g(t,y)&=&{{3\alpha^2(3w+2)}\over{32K}}\left[(1-Kb_0|y|)^{-{{16\kappa^2_5}
\over{9\alpha^2}}}-1\right].
\label{deffg}
\end{eqnarray}
This approximate solution reduces to the one obtained in Ref. \cite{cgrt} 
in the non-dilatonic limit $\alpha\to 0$.  

Just like the RS1 model \cite{rs1}, one can introduce another brane at some 
fixed distance from the original brane.  We put this second brane at 
$y=1/2$.  We denote the energy density and pressure of the matter fields on 
the brane at $y=1/2$ as $\varrho_*$ and $\wp_*$.  The tension of the brane 
at $y=1/2$ is denoted as $\sigma_*$.  Then, the equations of motion 
(\ref{expeineqs1}-\ref{eqndil}) have additional $\delta$-function terms 
$\sim\delta(y-1/2)$ associated with the additional brane and brane matter 
at $y=1/2$.   The metric components and the dilaton satisfy additional 
boundary conditions of the form (\ref{bc1}-\ref{dilbc}) at $y=1/2$ but 
with the respective quantities corresponding to those at $y=1/2$.  
Just as in the non-dilatonic brane case in the previous works \cite{cgkt,cgs} 
and as was pointed out \cite{bdl} to be a generic topological constraint for 
a model with compact extra dimension and two branes, the requirement of the 
stable radius (i.e. $b_0=const$) without a stabilizing potential restricts 
$\varrho$ and $\varrho_*$ to be related to one another in the following way:
\begin{equation}
\varrho_*=-{\cal W}_{1/2}\varrho,
\label{denrel}
\end{equation}
where ${\cal W}_{1/2}\equiv(1-Kb_0/2)^{{16\kappa^2_5}\over{9\alpha^2}}$ 
with time-dependent $K={{3\alpha^2}\over{16}}e^{-\alpha\phi_0(t)}\sigma$ 
(thereby, the ratio $\varrho_*/\varrho$ in general changes with time).  
[This constraint is understood \cite{cgrt} as a fine-tuning of the matter 
energy densities on the two branes required to maintain the constant radius 
$b$ of the extra dimension even in the absence of the stabilizing 
potential.]  As a consequence, we have unphysical result that the matter 
energy density on either of the branes has to be negative.  Since 
$\sigma_*<0$, the Friedmann equations on the second brane has the 
opposite sign from those of the conventional cosmology.  These problems 
can be resolved by introducing a radion potential $U(b)$, which 
stabilizes the radius $b$ of the extra dimension, as was explicitly 
shown in Ref. \cite{cgrt}.  The energy momentum tensor (\ref{emtens}) 
receives the following additional contribution from the radion potential:
\begin{equation}
T^{rad}_{00}=-n^2U(b),\ \ \ \ \ 
T^{rad}_{ij}=a^2U(b)\gamma_{ij},\ \ \ \ \ 
T^{rad}_{44}=b^2\left[U(b)+bU^{\prime}(b)\right].
\label{rademten}
\end{equation}
Assuming that the radion mass $m_{rad}$ is very heavy and the radion potential 
is approximated to $U(b)\approx M^5_b\left({{b-b_0}\over b_0}\right)^2$ (with 
$M^5_b\propto m_{rad}$ assumed to be the largest mass scale of the theory) 
near its minimum, one has $b=b_0$ from the (4,4)-component Einstein's 
equation without constraining the matter energy densities on the two branes.  
Then, the remaining components of the Einstein's equations averaged 
over the bulk (by integrating w.r.t. $y$) lead to the conventional 
Friedmann equations.  Note, however that since we averaged the Einstein's 
equations over $y$, the effective Friedmann equations contain contributions 
from matter fields on both branes, unlike the cosmological solutions of other 
related works and the solutions we discussed in the above, for which the 
effective Friedmann equations are local in $y$.  

Finally, we comment on the case in which the matter fields on the 
three-brane couple to the dilaton field, i.e. the $\delta{\cal L}_{mat}/
\delta\phi\neq 0$ case.  If we assume that there exists the frame in 
which the matter fields decouple from the dilaton field (just like the Jordan 
frame of the Brans-Dicke theory), namely the matter fields on the three-brane 
are minimally coupled with respect to a Weyl rescaled metric 
$\tilde{g}_{\mu\nu}=\Omega^2(\phi)g_{\mu\nu}$, then the dilaton equation 
(\ref{eqndil}) is modified to
\begin{eqnarray}
{8\over 3}{1\over n^2}\left[\ddot{\phi}-\dot{\phi}\left({\dot{n}\over n}
-3{\dot{a}\over a}-{\dot{b}\over b}\right)\right]-{8\over 3}{1\over b^2}
\left[\phi^{\prime\prime}+\phi^{\prime}\left({n^{\prime}\over n}
+3{a^{\prime}\over a}-{b^{\prime}\over b}\right)\right]
\cr
=2\alpha\Lambda e^{-2\alpha\phi}+\alpha\sigma e^{-\alpha\phi}
{{\delta(y)}\over b}+{1\over\Omega}{{d\Omega}\over{d\phi}}
T^{mat\,\mu}_{\ \ \ \ \ \mu}{{\delta(y)}\over b}.\ \ \ \ \ \ \ \ \ \ \ \ 
\label{eqndilmod}
\end{eqnarray}
So, the boundary condition (\ref{dilbc}) on the first derivative of $\phi$ 
at $y=0$ is modified to
\begin{equation}
{{[\phi^{\prime}]_0}\over{b_0}}=-{3\over 8}\alpha\sigma e^{-\alpha\phi_0}
-{3\over 8}(3\wp-\varrho){1\over\Omega_0}{{d\Omega_0}\over{d\phi_0}},
\label{dilbcmod}
\end{equation}
where $\Omega_0\equiv\Omega(\phi_0)$.  Note, the second term on the RHS 
vanishes for the radiation dominated universe, for which $\wp=\varrho/3$.  
So, Eq. (\ref{friedeq}) is modified to
\begin{eqnarray}
{\dot{a}^2_0\over a^2_0}+{\ddot{a}_0\over a_0}+{k\over a^2_0}=
\left({\kappa^4_5\over{36}}-{\kappa^2_5\over{32}}\alpha{1\over\Omega_0}
{{d\Omega_0}\over{d\phi_0}}\right)(\varrho-3\wp)\sigma e^{-\alpha\phi_0}
+\left({\kappa^4_5\over{18}}\sigma^2-{\kappa^2_5\over{64}}
\alpha^2\sigma^2\right.
\cr
\left.+{\kappa^2_5\over 3}\Lambda\right)e^{-2\alpha\phi_0}
-{\kappa^4_5\over{36}}\varrho(\varrho+3\wp)
-{\kappa^2_5\over{64}}(3\wp-\varrho)^2{1\over\Omega^2_0}\left({{d\Omega_0}
\over{d\phi_0}}\right)^2-{4\over 9}\kappa^2_5\dot{\phi}^2_0.
\label{modfriedeq}
\end{eqnarray}
In the limit $\varrho\ll\sigma$, the conventional Friedmann equations 
with $H^2\sim\varrho$ behavior are reproduced to the leading order, 
however the coefficient is modified due to the coupling of the matter 
fields to $\phi$.  On the positive tension brane, in order for the 
Friedmann equations with the correct sign to be reproduced, $\alpha$, 
$\phi_0$ and $\Omega_0$ are constrained to satisfy $\alpha\Omega^{-1}_0
d\Omega_0/d\phi_0<8\kappa^2_5/9$.  Particularly when $\Omega=e^{\beta\phi}$ 
with a constant $\beta$, this constraint restricts the allowed values of 
$\alpha$ and $\beta$ to satisfy $\alpha\beta<8\kappa^2_5/9$.  The coupling 
of the matter fields to $\phi$ also induces another subleading correction 
(the second to the last term on the RHS) to the Friedmann equations.

\end{document}